\documentstyle[pre,aps]{revtex}
\begin{document}
\small
\setcounter{page}{0}
\input epsf
\title{Retrieval behavior and thermodynamic properties of symmetrically diluted Q-Ising neural networks}
\author{W. K. Theumann and R. Erichsen Jr.}
\address{Instituto de F{\'\i}sica,
Universidade Federal do Rio Grande do Sul,\\ 
Caixa Postal 15051. 91501-970 Porto Alegre, RS, Brazil}

\date{\today}
\maketitle

\thispagestyle{empty}
\begin{abstract}
The retrieval behavior and thermodynamic properties of symmetrically diluted Q-Ising neural networks are derived and studied in replica-symmetric mean-field theory generalizing earlier works on either the fully connected or the symmetrical extremely diluted network. Capacity-gain parameter phase diagrams are obtained for the $Q=3$, $Q=4$ and $Q=\infty$ state networks with uniformly distributed patterns of low activity in order to search for the effects of a gradual dilution of the synapses. It is shown that enlarged regions of continuous changeover into a region of optimal performance are obtained for finite stochastic noise and small but finite connectivity. The de Almeida-Thouless lines of stability are obtained for arbitrary connectivity, and the resulting phase diagrams are used to draw conclusions on the behavior of symmetrically diluted networks with other pattern distributions of either high or low activity.

\end{abstract}


\setcounter{page}{1}
\label{sec:level1}


\section{Introduction}

The storage and retrieval properties, as well as the dynamics, of large attractor neural networks have been studied over some time and numerous results are now available. The presence of an exponentially large number of unwanted spin-glass-like states in fully connected networks may limit severely the storage capacity, the information content and the performance, in particular the retrieval quality of previously stored patterns which act as attractors in the dynamics of the network. Also, the size of the corresponding basins of attraction may be considerably reduced. Other tasks, as the categorization or generalization ability of a network, are also impaired due to the presence of such states and, except for a low storage ratio, that is for a reduced level of stochastic noise, a performing network is very likely to be trapped in one of those states preventing  the occurence of finite overlaps with the patterns of interest. 
   
The study of the equilibrium behavior of the Hopfield model with binary neurons and extremely low {\it symmetric} connectivity $c$ suggested that \cite{WS91}, except for zero storage ratio $\alpha$ or in the absence of synaptic noise $T$, the properties of the model may be quite different from those of the extremely diluted network with asymmetric synapses, which has a trivial dynamics \cite{DGZ87}. In particular, the equations for the order parameters are equivalent to those of the Sherrington and Kirkpatrick (SK) spin-glass (SG) model \cite{SK75}. 

It has been shown, nevertheless, that strong symmetric dilution of synapses in the Hopfield model \cite{H82,HKP91} may considerably reduce the stability of spin-glass states. Indeed, in the limit $c\rightarrow 0$, the retrieval states are globally stable in their whole domain of existence of the $(\alpha, T)$ phase diagram, thus leaving only locally stable SG states \cite{CN92}, in contrast to the situation in the fully connected Hopfield model \cite{AGS87}. It has also been shown that already a gradual dilution reduces the stability of the SG states in the region where they compete with the retrieval states and that the storage capacity of the network is increased \cite{CN92}. These are important results on the performance of a network which also suggest that the basins of attraction of the memory (retrieval) states should be increased in symmetrically diluted networks.

Networks of Q-Ising neurons in various architectures have been studied over some time, in particular, fully connected, layered feed-forward and extremely dilute networks with asymmetric connectivity $[8-15]$. They are of interest for possible hardware applications and an eventual biological modeling, because of the increased flexibility of the states of the network to account for complex neural behavior. It has been suggested that an attractor neural network model with multi-state neurons may describe the short term memorization performance of the $CA_3$ region of the hyppocampus in both the brain of primates and in the human brain, and results of numerical simulations for the selective performance of a network with asymmetric synapses and small connectivity ($c=0.2$) are now available \cite{RTFP97}. However, complete phase diagrams that give a global picture of the performance of a network and a physical explanation of why biological networks seem to prefer a low connectivity are still missing. Attractor network models with a Hebbian type of learning rule may also serve to account for long-term memory in the brain \cite{Ar95}. 

Both the parallel dynamics and the equilibrium properties of extremely diluted networks with symmetric connectivity between multi-state Q-Ising neurons have been studied in recent works \cite{BJS99,BCS00}. The symmetry of the synaptic connections allows for a detailed balance assumption in the dynamics and enables one to perform full analytic calculations of the equilibrium properties of a network. This is particularly convenient in the search for phase diagrams. So far, there are no results, to our knowledge, concerning the memorization performance of symmetrically diluted Q-Ising networks with small but finite connectivity and low-activity patterns, which characterize a biological network. 

The purpose of the present work is to consider the retrieval performance of an attractor neural network modeled on some of the features of a biological network. We study here the equilibrium behavior of a symmetrically diluted Hopfield model with finite connectivity and low-activity units in Q-Ising states. Multi-state networks are known to have complex properties and our main aim here is to find out how that behavior is changed by a finite dilution of the synapses and the low activity of the embedded patterns. We consider explicitly a $Q=3$, $Q=4$ and a continuous $Q=\infty$-state network.

The outline of the paper is the following. In Sec. II we present the model and show how the Hamiltonian becomes the sum of an effective Hopfield-model Hamiltonian and that of a Sherrington-Kirkpatrick-like spin-glass model. The replica-symmetric mean-field theory for the model and the relevant order parameters are derived in Sec. III and explicit expressions are presented in the Appendix. The results for the phase diagrams and the retrieval performance are discussed in Sec. IV, and our conclusions with a summary of the results can be found in Sec. V. 

\section{The model}

Attractor neural networks are dynamical systems. Consider a network of $N$ neurons, $i=1,\dots,N$. At the time step $t$, the state of neuron $i$ is described by the variable $S_i(t)$, that can be in any one of the $Q$ Ising states
\begin{equation}
\sigma_k=-1+\frac{2(k-1)}{Q-1}
\label{1}
\end{equation}
in the interval [-1,1], for $k=1,\dots,Q$. A macroscopic set of $p$ independent and identically distributed random patterns $\{\xi_i^{\mu};
\mu=1,\dots,p; i=1,\dots,N\}$, with $p={\alpha}c N$, is embedded in the network by means of a Hebbian-like learning rule, specified below. Here, ${\alpha}$ is the storage ratio per connected site and $c$ is the connectivity of the network. Every bit of each pattern can be in any one of $Q$ equally spaced states, also in the interval [-1,1], which are assumed to have zero mean and variance 
\begin{equation}
\left\langle(\xi_i^{\mu})^2\right\rangle=a\leq 1. 
\label{2}
\end{equation}
This is a measure of the size of the patterns, describing their mean activity, that accounts for bits that either are turned off or depressed. 

The first task to be performed by the network is to attain a finite storage ratio $\alpha$ and this ratio can be optimized for patterns of a given size by means of an appropriate tuning of the states of the network, discussed below. The second aim is to reach a sufficiently small Hamming distance  
\begin{equation}
d_{H}({\bf{\xi}}^{\mu},\sigma)=N^{-1}\sum_{i}(\xi_i^{\mu}-S_{i})^{2}            \label{3}
\end{equation}
between the network state $\{S_i\}=(S_i,\dots S_N)$ and a given pattern $\{\xi_i^{\mu}\}$, as a measure of the retrieval performance of the network. 
This depends on both the overlap 
\begin{equation}
m^{\mu}=(aN)^{-1}\sum_{i}\xi_i^{\mu}S_{i}
\label{4}
\end{equation}
and the dynamical activity of the network, $a_D=(aN)^{-1}\sum_{i}S_i^{2}$. 

We consider next the dynamics of the partially connected model. For a given configuration $\{S_i\}$ of the network, the local field acting on neuron $i$ is given by
\begin{equation}
h_i(\{S_i\})=\sum_{j}J^{d}_{ij}S_j\;,
\label{5}
\end{equation}
where $J^{d}_{ij}$ is the synaptic coupling between neurons $i$ and $j$. We assume that the synapses of the network are symmetrically diluted being left with a {\it finite} connectivity $c$ (the fraction of connected neurons) which, eventually, may become vanishingly small. Specifically, the synapses are of the Hebbian-like form,
\begin{equation}
J^{d}_{ij}=\frac{c_{ij}}{acN}\sum_{\mu=1}^p\xi_i^{\mu}\xi_j^{\mu}\,\
\label{6}
\end{equation}
for our non-sparse network, in which $\left\langle(\xi_i^{\mu})\right\rangle=0$. Here, $\{c_{ij}=c_{ji};i,j=1,\dots,N\}$ is a set of independent identically distributed random variables, such that $c_{ij}=1$ with probability $c$ and zero with probability $1-c$, while $c_{ii}=0$. Thus, the symmetric dilution introduces an additional randomness both into the dynamics of the network, which becomes non-trivial even in the extremely diluted case due to feedback loops \cite{BJS99}, as well as in the thermodynamics. We are interested, in the following, in the case of a large connected network, in which $cN$ is very large.

The state of each neuron is updated asynchronously according to a Glauber (single spin-flip) dynamics \cite{HKP91} in which the transition probabilities are given by 
\begin{equation}
P(S_j(t+\Delta t)=\sigma_k|\{S_i(t)\})
	\quad=\frac{\exp[\beta\epsilon_j(\sigma_k|h_j(\{S_i(t)\}))]}
	{\sum_{l=1}^Q\exp[\beta\epsilon_j(\sigma_l|h_j(\{S_i(t)\}))]}\;,
\label{7}
\end{equation}
where $\beta=1/T$ is the inverse synaptic noise (temperature) and the single site energy, $\epsilon_j(s|h)$, is given by
\begin{equation}
\epsilon_j(s|h)=-hs+\theta s^2\,.
\label{8}
\end{equation}
Here, $\theta$ is a tuning parameter favoring local states of small 
dynamical activity. In the absence of stochastic noise, the deterministic
evolution of the system follows the updating rule
\begin{equation}
S_j(t+\Delta t)=\Theta_{dyn}(h_j(t))\,,
\label{9}
\end{equation}
where $\Theta_{dyn}(x)$ is the non-decreasing step function, for finite $Q$,
\begin{equation}
\Theta_{dyn}(x)=\sum_{k=1}^Q\sigma_k[\Theta(\theta(\sigma_{k+1}+\sigma_k)-x)
		-\Theta(\theta(\sigma_k+\sigma_{k-1})-x)]
\label{10}
\end{equation}
with $\sigma_0=-\infty$ and $\sigma_{Q+1}=+\infty$, in which $\Theta(x)=1$, if
$x\geq 0$, and zero otherwise. Clearly, $\theta$ is a threshold parameter since the state of neuron $j$ assumes the value $\sigma_k$
given by Eq.~(\ref{1}) if the local field $h_j$ is bound by
$\sigma_k+\sigma_{k-1}\leq h_j/\theta\leq\sigma_k+\sigma_{k+1}$. The width of
the intermediate states with constant $\sigma_k$ for $1<k<Q$ (that is,
excluding the limiting values of $\sigma_k=\pm 1$), is given by
$4\theta/(Q-1)$. Thus, the width of the zero state for the three-state network
studied below is $2\theta$. In the limit $Q\rightarrow\infty$, the
input-output function becomes the piecewise linear function
\begin{equation}
\Theta_{dyn}(x)={\rm sign}(x)
	{\rm min}\left(\frac{\left| x\right|}{2\theta}, 1\right)\,,
\label{11}
\end{equation}
where ${\rm min}(x,y)$ means the minimum between $x$ and $y$. The slope of the
linear part, $1/2\theta$, is the gain parameter of the continuous network.

Both, the locally stable states of the dynamics, as well as the globally stable states that characterize the equilibrium thermodynamic properties of the diluted
network, that follow from the above dynamics, are described by the Hamiltonian
\begin{equation}
H=-\sum_{(i,j)}J^{d}_{ij}S_iS_j+\theta\sum_{i}S_i^2\;,
\label{12}
\end{equation}
where the first sum is over all distinct pairs of neurons $(i,j)$. Eventually, a field-dependent term, $h_1\sum_i\xi^{1}_iS_i$, may be added to generate the overlap with a specific pattern, and this will be implicitly assumed below.

We adapt next the procedure of Viana and Bray \cite{VB85} for diluted spin glasses
in order to deal with the random dilution. In distinction to their case, which is that of a strongly interacting spin glass, we have here a diluted network with randomness in the patterns and weak interactions between units. The latter allows for an exact truncation to the relevant order in $1/cN$ that is sufficient for the mean-field calculation of the following section.

Consider the disorder dependent part of the Hamiltonian in the exponentiated form 
\begin{equation}
G_{\beta}(\{S_i\})
	=\exp\{{\beta}\sum_{(i,j)}J^{d}_{ij}S_iS_j\}\,.
\label{13}
\end{equation}
Given a fixed set $\{\xi^{\mu}_i\}$ of patterns embedded in the network, we first have to build up a finite connectivity between units. That is to say, we have to find a network such that the mean of $c_{ij}$ is precisely $c$ and, to this end, we have to perform first an average over these random variables. The average
over the random patterns, that is necessary to evaluate the performance of the network, comes at a later stage. The configurational average over the set $\{c_{ij}\}$ of the $n$ times replicated function $G_{\beta}(\{S_i\})$ becomes 
\begin{equation}
\left\langle G^{n}_{\beta}(\{S_i\})\right\rangle_c=
\exp\{\sum_{(i,j)}\ln(1+c(\exp({\beta}J_{ij}
\sum_{\alpha}S^{\alpha}_iS^{\alpha}_j)-1))\}\,,
\label{14}
\end{equation}
where $J_{ij}$ denotes the value of $J^d_{ij}$ for $c_{ij}=1$. We are interested in dense networks, that is, $c=O(1)$ and large $cN$, letting eventually $c \rightarrow 0$ after the thermodynamic limit. For finite $\alpha=p/cN$, the weak $J_{ij}=O(\sqrt {\alpha/cN})$ may be used to expand the logarithm to next-to-leading order in $J_{ij}$. Using the law of large numbers to deal with $(J_{ij})^{2}$ we obtain 
\begin{equation}
\left\langle G^{n}_{\beta}(\{S_i\})\right\rangle_c=\left\langle \exp\{\beta\sum_{(i,j)}J^{eff}_{ij}S^{\alpha}_iS^{\alpha}_j\}\right
\rangle_{\delta}\,,
\label{15}
\end{equation}
in which the effective coupling $J^{eff}_{ij}$ is given by
\begin{equation}
J^{eff}_{ij}=\frac{1}{aN}\sum_{\mu=1}^p\xi_i^{\mu}\xi_j^{\mu}+\delta_{ij}\,. 
\label{16}
\end{equation}
The first term is the Hebbian synapsis of the fully connected network with multi-state patterns of activity $a$ and $\delta_{ij}$ is a Gaussian random variable with zero mean and a pattern-independent variance $\Delta^{2}/N=\alpha(1-c)/N$; the brackets on the right-hand side of Eq.(15) denote an average over this random variable. This generalizes an earlier result by Sompolinsky \cite{S87} showing that dilution appears as a synaptic noise. 

Thus, the symmetric dilution introduces an effective Hamiltonian, in the large $cN$ limit,
\begin{equation}
H^{eff}=-\sum_{(i,j)}J^{eff}_{ij}S_iS_j+\theta\sum_{i}S_i^2\;,
\label{17}
\end{equation}
where the first term is a sum of a Hopfield-model Hamiltonian and a kind of
SK-model term. This is used in the following section to derive the mean-field theory for the model.

\section{Mean-field Theory}

We consider now the mean-field theory for finite $\alpha$ and for any connectivity $c=O(1)$. The averaged free energy per connected site is given by
\begin{equation}
f(\beta)=-\lim_{cN\rightarrow\infty}\frac{1}{\beta cN}\left\langle
	\left\langle\ln Z_{eff}(\beta)\right\rangle_{\{\delta_{ij}\}}
	\right\rangle_{\{\xi^{\mu}\}}\,,
\label{18}
\end{equation}
with the averages first over the Gaussian noise in the synaptic interactions and then over the pattern distribution. The effective partition function is then given by
\begin{equation}
Z_{eff}(\beta)=\sum_{\{S_i\}}\exp(-\beta H^{eff})\,,
\label{19}
\end{equation}
and this is used in the replica method to perform the random averages by means of
\begin{equation}
\left\langle\left\langle\ln Z_{eff}(\beta)
	\right\rangle_{\{\delta_{ij}\}}
	\right\rangle_{\{\xi^{\mu}\}}=\lim_{n\rightarrow 0}\frac{1}{n}
	\left(\left\langle\left\langle Z^{n}_{eff}(\beta)
	\right\rangle_{\{\delta_{ij}\}}\right\rangle_{\{\xi^{\mu}\}}
	-1\right)\,.
\label{20}
\end{equation}
The essential point is then the calculation of the averaged replicated partition function. Assuming, as usual, that a finite number of patterns $\xi^{\nu}_i$ is condensed, with finite overlaps with the state of the network, we perform the random noise average and sum over the uncondensed patterns to obtain
\begin{eqnarray}
&&\!\!
\left\langle\left\langle Z^{n}_{eff}(\beta)
	\right\rangle_{\{\delta_{ij}\}}
	\right\rangle_{\{\xi^{\mu}\}}
        =\sum_{\{S^{\alpha}_i\}}
        \exp\{\frac{\Delta^{2}\beta^{2}}{N}
         \sum_{\alpha,\beta}(\sum_iS^{\alpha}_iS^{\beta}_i)^2\}\nonumber\\
	&&\quad\times
        \int\prod_{\alpha}\frac{\sqrt{\beta N}{\rm d}m^{\nu}_{\alpha}}
	{\sqrt{2\pi}}\exp\{-\frac{\beta N}{2}
	\sum_{\alpha}(m^{\nu}_{\alpha})^2)+np\beta F\}\nonumber\\
	&&\quad\times
	\left\langle\exp\{\beta\sum_{i,\alpha}\left[
	m^{\nu}_{\alpha}\xi^{\nu}_iS^{\alpha}_i
	-\theta(S^{\alpha}_i)^2\right]\}
        \right\rangle_{\{\xi^{\mu}_i\}}\,.
\label{21}
\end{eqnarray}
where, the sum over $\{S^{\alpha}_i\}$ counts the configurations for all possible replicas $\alpha$. The dilution appears only in the first factor and
\begin{equation}
\exp({\beta}npF)=\prod_{\mu>1}\left\langle\exp\{\frac{\beta}
	{2N}\sum_{i\neq j}\sum_{\alpha}S^{\alpha}_iS^{\alpha}_j
	\xi^{\mu}_i\xi^{\mu}_j\}\right\rangle_{\{\xi^{\mu}_i\}}
\label{22}
\end{equation}
is the average over the uncondensed patterns.

For both, the calculation of the function $F$ and the linearization of the quadratic form in the dilution term, we introduce the replica matrix elements
\begin{equation}
q_{\alpha\beta}=Q_{\alpha\beta}\equiv\frac{1}{N}\sum_iS^{\alpha}_i
S^{\beta}_i\;\;\;\;{\rm if}\;\;\alpha\neq \beta\,,
\label{23}
\end{equation}
of the SG order parameter and
\begin{equation}
\tilde{q}_{\alpha}=Q_{\alpha\alpha}\,,
\label{24}
\end{equation}
which is the dynamical activity $a_D$ of the network. The latter is one only in the case of binary units and, in general, $q_{\alpha\beta}\leq \tilde{q}_{\alpha}\leq 1$. In the thermodynamic limit $N\rightarrow\infty$, we obtain \cite{IT92} 
\begin{equation}
n\beta G=-\frac{1}{2}{\rm tr}\ln\left(1-\beta{\bf q}\right)
	-\frac{1}{2}\beta\,{\rm tr}\,{\bf q}\,.
\label{25}
\end{equation}
where ${\bf q}$ is the matrix of elements $q_{\alpha\beta}$ and $\tilde{q}_{\alpha}$. Introducing as usual the overlap parameter $r_{\alpha\beta}$ associated to the correlation between the overlaps of the patterns that do not condense, we restrict ourselves to replica symmetry, in which
\begin{eqnarray}
        m_{\nu}&=&m^{\nu}_{\alpha}\\	
        q&=&q_{\alpha\beta}\\
	a_D&=&\tilde{q}_{\alpha}\\
	r&=&r_{\alpha\beta}.
\label{26}
\end{eqnarray}

The free-energy per connected site, in the thermodynamic $N\rightarrow\infty$ limit, then follows as  
\begin{eqnarray}
f(\beta)&=&\frac{a}{2}\sum_{\nu}(m_{\nu})^{2}+\frac{{\alpha}c}{2\beta}
\left[\ln(1-\chi)+\frac{\chi}{1-\chi}+\frac{q\beta\chi}
{(1-\chi)^2}\right]\nonumber\\
&+&\alpha(1-c)(\frac{\chi^2}{4\beta}+\frac{q\chi}{2})
-\frac{1}{\beta}\left\langle
\int{\cal D}z\,\ln\sum_{\{S\}}\exp(\beta{\cal H}_{eff})\right\rangle_{\{\xi^{\nu}\}}\;,
\label{27}
\end{eqnarray}
where ${\cal D}z={\rm d}z\exp(-z^2/2)/\sqrt{2\pi}$ is a Gaussian measure, and $\chi=\beta(a_D-q)$, given by $\beta\sum_i(\langle S_i^2\rangle-\langle S_i\rangle^2)/N$, is the susceptibility of the network. The new, site independent effective Hamiltonian ${\cal H}_{eff}$, is given by
\begin{equation}
{\cal H}_{eff}=S\left(\sum_{\nu}m_{\nu}\xi^{\nu}
	+\sqrt{{\alpha}rc}z-\tilde{\theta}S\right)\,,
\label{28}
\end{equation}
in terms of which the thermal averages are defined as
\begin{equation}
\langle S^n(z)\rangle=\frac{\sum_{\{S\}}S^{n}\exp(\beta{\cal H}_{eff})}
	{\sum_{\{S\}}\exp(\beta{\cal H}_{eff})}\,.
\label{29}
\end{equation}
Note, incidentally, that the explicit term in $\alpha(1-c)$ in the free energy comes from the variance of the Gaussian random noise due to the dilution.

The order parameters that describe the performance of the network are given by the saddle-point equations
\begin{equation}
m_{\nu}=\frac{1}{a}\left\langle\xi^{\nu}\int{\cal 
	D}z\,\langle S(z)\rangle\right\rangle_{\{\xi^{\nu}\}}\,,
\label{30}
\end{equation}
\begin{equation}
q=\left\langle\int{\cal D}z\,
	\langle S(z)\rangle^2
	\right\rangle_{\{\xi^{\nu}\}}\,,
\label{31}
\end{equation}
and the susceptibility becomes
\begin{equation}
\chi=\frac{1}{\sqrt{\alpha rc}}\left\langle\int{\cal D}z\,
	z\langle S(z)\rangle
	\right\rangle_{\{\xi^{\nu}\}}\,,
\label{32}
\end{equation}
in which $\langle S(z)\rangle$, for a given Gaussian noise $z$ is given by Eq.(32). Noting that the effective Hamiltonian ${\cal H}_{eff}$ is formally the same as either that of
the fully connected or the extremely diluted network, with a different
stochastic Gaussian noise and different effective threshold $\tilde\theta$,
for each connectivity $c$, our equations for $m_{\nu}$, $q$ and $\chi$
will be formally similar to the equations for both these networks, and explicit expressions are given in the Appendix.

The parameter $r$ follows from the algebraic saddle-point equation
\begin{equation}
r=q(\frac{1}{(1-\chi)^2}+\frac{1-c}{c})\,,
\label{33}
\end{equation}
and
\begin{equation}
\tilde{\theta}=\theta-\frac{\alpha\chi}{2}(1+\frac{c\chi}{1-\chi})\,,
\label{34}
\end{equation}
is the effective width of the intermediate states. Eventually, depending on the state of the network specified by
the dynamical activity 
$a_D$ and the SG order parameter $q$, $\tilde{\theta}$
may become negative, favoring an order with large absolute values for $S$.
Whenever this is the case the network acts, at zero temperature, as a binary network with zero threshold, as will be seen below. The combination 
$\sum_{\nu}m_{\nu}\xi^{\nu}+\sqrt{{\alpha}rc}z$, in which the second term is the Gaussian noise due to the macroscopic number of uncondensed patterns, can be viewed as an effective random field that will influence the network performance through a competition with the effective threshold $\tilde{\theta}$. 

Having performed the thermodynamic limit, one may now allow {\it any} value for the connectivity within $0\leq c\leq 1$, including the limit $c\rightarrow 0$. It can easily be seen that one recovers for $f(\beta)$ both the mean-field free energy for the fully connected network when $c=1$ \cite{BRS94} and that for the extremely dilute case, when $c \rightarrow 0$ \cite{BCS00}. In this limit, $rc=q$ and $\tilde{\theta}=\theta-\alpha{\chi}/2$, whereas for $c=1$ we have, $r=q/(1-\chi)^2$ and $\tilde{\theta}=\theta-\alpha{\chi}/2(1-\chi)$, also in agreement with the known results. These relations are valid for any temperature parameter $1/\beta$. Thus, we have the complete form for the replica-symmetric free energy for arbitrary connectivity.

Note, incidentally, that $rc=q$ not only in the above limit but also for $\alpha=0$,
in the $\beta \rightarrow\infty$ limit, for any connectivity $c$, (that is, for any architecture) and for all $Q$, and this is based on $\chi \rightarrow 0$ in that limit. Otherwise, the susceptibility remains finite, even at zero temperature, since $q\rightarrow a_D$, when $\beta\rightarrow\infty$, 
while at finite temperature we have, in general, $q\leq a_D$.

The limit of stability of the replica-symmetric solution comes from the 
study of quadratic fluctuations of the free-energy in the vicinity of the
symmetric saddle-point. Following the de Almeida and Thouless (AT) 
analysis \cite{AT78}, we obtain 
\begin{equation}
\frac{{\alpha}rc\beta^2}{q}\left\langle\int{\cal D}z\,
        \left[\langle S^2(z)\rangle-\langle S(z)\rangle^2\right]^2
	\right\rangle_{\{\xi^{\nu}\}}
        \leq 1\,,
\label{35}
\end{equation}
as the stability condition for the replica-symmetric solution. This equation is to be solved together with the saddle-point equations for the order parameters.

The formal results obtained so far are valid for any finite number of condensed patterns with finite overlaps with the state of the network. We are mainly interested in this work in the retrieval performance with a single condensed pattern, and this will be discussed in the following section. 

\section{Retrieval and Thermodynamic Properties}
 
For the retrieval of a single condensed pattern, say $\xi^1$, we have
$m_{\nu}=m\delta_{{\nu}1}$, and omit the index $1$ from now on. We consider separately the results for $Q=3$, $Q=4$ and $Q=\infty$ that follow by solving the saddle-point equations and restrict ourselves to the simplest case of uniformly distributed patterns. 

\subsection{The three-state network}

In the $Q=3$ case, the patterns take the values 
$\pm 1$ with probability $a/2$ and $0$ with probability $(1-a)$, in which $a=2/3$ in the case of uniformly distributed patterns. The effective transfer function $S_{\beta}(h,\tilde\theta)=\langle S(z)\rangle$ that follows from the average in Eq.(32) is given by
\begin{equation}
S_{\beta}(h,\tilde\theta)=\frac{\sinh(\beta h)}{\frac{1}{2}\exp(\beta\tilde\theta)+\cosh(\beta h)}
\label{36}
\end{equation}
which becomes
\begin{equation}
S_{\infty}(h_s,\tilde\theta)\equiv\lim_{\beta\rightarrow\infty}
S_{\beta}(h_s,\tilde\theta)={\rm sgn}(h_s)\Theta(|h_s|-\tilde\theta)\,,
\label{37}
\end{equation}
in the zero temperature limit. Clearly, when $\tilde\theta<0$, the network acts as a binary network at $T=0$.

In the cases of the fully connected \cite{BRS94} or the extreme symmetrically  diluted network \cite{BCS00}, explicit closed form expressions that signal the appearance of either a retrieval or a spin-glass phase, at zero temperature, have been obtained in the $\tilde\theta \rightarrow 0$ limit when $\tilde\theta \leq 0$. These are particularly useful to understand the low-threshold behavior of the phase diagrams. In the present case, of a network with finite and less than complete connectivity, closed form expressions for the onset of the ordered phases cannot be obtained and one has to resort fully to numerical solutions.

Nevertheless, the zero-temperature behavior of the thermodynamic transition can easily be analyzed as in previous works, to demonstrate that the retrieval state corresponds to the most stable phase, despite the presence of a spin-glass and a paramagnetic phase (which is a frozen zero-spin state). Indeed, the physical free energy, $f=-f(\beta)$ at zero temperature becomes
\begin{equation}
f=-\frac{a}{2}m^{2}-\frac{\alpha}{2}\chi rc+\tilde\theta rc\,.
\label{38}
\end{equation}
Since the susceptibility vanishes in the $\alpha \rightarrow 0$ limit when $T=0$ for any of the three phases, and at the same time $rc \rightarrow q$, the retrieval free energy is the minimum whenever $\theta \leq 1/2$. Note that the susceptibility of the paramagnetic phase,
\begin{equation}
\chi=\frac{2\beta}{\exp(\beta \tilde\theta)+2}\,,
\label{39}
\end{equation}
also vanishes in the zero temperature limit for finite $\alpha$ and $\tilde\theta$ converging to $\theta$, ensuring a minimum
retrieval free energy for small $\alpha$.

Although much emphasis is often made on the thermodynamic transition to globally stable retrieval states, which have the lowest free energy, it is worth keeping in mind that neural networks are dynamical systems with accessible {\it locally} stable retrieval states, in particular in the presence 
of some amount of noise. As far as the performance of the network is concerned, these are the most interesting states and they usually appear for higher values of $\alpha$ \cite{HKP91}.

We are interested here in the characteristic features of the phase diagrams and the specific performance of the network. To see the effects of a gradual change in the connectivity, we show in Fig. 1 the $(\alpha,\theta)$ phase diagrams for $T=0$. The full lines represent the phase boundaries where the locally stable retrieval states appear at the critical storage ratio $\alpha_c$, whereas the long-dashed lines indicate the thermodynamic transitions to the globally stable retrieval phase. The SG phase appears to the left of the short-dashed lines representing the boundary to the paramagnetic state. To distinguish in what follows the transitions involving locally stable states from the thermodynamic transitions, we refer to the former simply as retrieval transitions. As long as the connectivity remains finite, all the transitions are discontinuous and, as usual in connected networks, the SG state is globally stable only above the thermodynamic transition \cite{AGS87}.

Consider first the case of the fully connected network, with $c=1$, which has 
been redone and completed here for the purpose of comparison. There are two retrieval regions, I and II for small $\alpha$, separated by a sharp phase boundary, and this is the case both below and above the thermodynamic transition. The first is a region of non-optimal performance characterized by a moderately large Hamming distance which decreases with increasing $\theta$, whereas in region II the Hamming distance is small, dropping discontinuously at the phase boundary between the two regions with optimal network performance along the dotted line.

The situation should change with decreasing connectivity, even at zero temperature.
Due to the synaptic noise produced by the dilution, given by the variance $\Delta^{2}/N=\alpha(1-c)/N$, one now expects an end to the discontinuous transition between the retrieval regions I and II
at a $c$-dependent critical point for finite $\alpha$, below $\alpha_c$. This starts to appear for $c\simeq 0.63$, with $\theta\simeq 0.325$ and $\alpha\simeq 0.0282$, as shown in the inset of the figure for the retrieval transition. A critical point also appears at the thermodynamic transition and in Fig. 1 we show the results for $c=0.5$ and $c=0.25$. There can be now a continuous transition with increasing $\theta$ induced by stochastic noise between the non-optimal and optimal performance retrieval states above the critical point suggesting the presence of increasingly larger regions of continuous changes. The interesting point is that this is a feature that already starts to appear for finite and {\it intermediate} connectivity, between that of the fully connected network and the $c\rightarrow 0$ limit.

To understand the role of the threshold one may use as a guide the case of vanishing stochastic noise, with $\alpha \rightarrow 0$. When $\theta$ is small, the state of a unit will be essentially $S_i=\pm 1$, with an overlap $m_I^{\mu}=(aN)^{-1}\sum_{i}|\xi_i^{\mu}|=1$ in region I for uniformly distributed three-state patterns $\xi_i^{\mu}$, that take the values $-1$, $0$ and $+1$. Since $q=1$, the Hamming distance in this region will be $d_{H}(I)=1/3$. On the other hand, as $\theta$ becomes larger, the state $S_i=0$ becomes increasingly important and, despite the fact that $S_i=\xi_i^{\mu}$ also yields an overlap $m_I^{\mu}=1$, the SG order parameter $q$ is now reduced to the activity $2/3$. Thus, the resulting Hamming distance $d_{H}(II)$ will be vanishingly small. Note that these results do not depend on the connectivity and they are, therefore, independent of the architecture of the network. This also follows from the zero-temperature saddle-point equations as can easily be checked with the Appendix. In the case of a finite non-zero stochastic noise, instead, the performance of the network becomes explicitly dependent on the connectivity, but the overall qualitative dependence on $\theta$ below the critical phase boundaries is expected to follow that at $\alpha=0$.

The critical storage capacity now increases with decreasing connectivity and the presence of two comparable maxima for $\alpha_c$ is only a feature of intermediate $c$. Indeed, as the synapses are further diluted a single maximum is left, albeit with a shift to higher values of $\theta$. Finally, it is also worth noting that, in the $c\rightarrow 0$ limit, the retrieval state is the globally stable phase everywhere below the critical $\alpha_c$ line and to the left of the globally stable paramagnetic phase, despite the relatively large stochastic noise due to the presence of spin-glass states in most of this region.

The zero-temperature results presented so far are not stable to replica-symmetry breaking perturbations but it is expected that most of the features described here will be present at already a small but finite temperature above the AT line, shown as dash-dotted in Fig. 2, for $\theta=0.2$. The full lines again represent the transitions to the locally stable retrieval states and the thermodynamic transitions are not shown. Note that, even for small connectivity, there is only a low synaptic-noise region in which the network is not stable to replica-symmetry breaking perturbations. For a larger $\theta=0.5$ we expect a similar behavior for the transition to the locally stable retrieval state with decreasing connectivity as that found before for the fully connected network, but still quite different from the behavior for lower $\theta$ \cite{BRS94}.  

\subsection{The four-state network}                                                                                                    
In the case of $Q=4$ the phase diagrams are more involved. The patterns are assumed to take the values $\pm 1$ with probability $\tilde a/2$ and $\pm 1/3$ with probability $(1-\tilde a)/2$, in which $\tilde a=(9a-1)/8$, where $a=5/9$ for uniformly distributed patterns. We consider in the following the zero-temperature behavior of the network and begin with the fully connected case as a guide. We recover precisely the retrieval phase boundaries found before \cite{BRS94}.

Similar results with exclusively sharp phase boundaries, between now enlarged ordered regions, are found for somewhat lower connectivity, as shown in Fig. 3(a) for $c=0.5$. For low $\alpha$, we find three different ordered retrieval ferromagnetic phases $FM_{3\Delta}$, $FM_1$ and $FM_{\Delta}$, in a previous notation \cite{BRS94} and characterized below, separated by discontinuous phase boundaries and in which $\Delta=2/5$ for uniformly distributed patterns. As $\alpha$ increases, these phases disappear discontinuously at the critical phase boundaries $\alpha_c$ into the $SG$ phase. The three phases correspond to possible locally stable states which become globally stable at the thermodynamic transition for lower $\alpha$, not shown in the figure for simplicity. Which of the locally stable states is actually reached in the dynamic evolution of the network will depend, as usual, on their basins of attraction and the choice of the initial state. 

One may consider four regimes in the $\alpha\rightarrow 0$ limit. First, when $\theta$ is small, that is within the phase $FM_{3\Delta}$, mainly the high activity states are favored with $S_i={\rm sgn}(\xi_i)$ with an overemphasized overlap $m=6/5$; the spin-glass order parameter and the Hamming distance become $q=1$ and $d_H=2/9$, respectively. For larger and intermediate $\theta$, there should be a phase, called $FM_1$, characterized by states of the network that follow essentially the patterns, with $S_i=\xi_i$ and an overlap given by $m=1$, the spin-glass order parameter $q=a=5/9$ and a vanishing Hamming distance. Between the small and the intermediate $\theta$ regimes there could be a coexistence region of the phase $FM_{3\Delta}$ with the phase $FM_1$ where the states of the network start to recognize the full structure of the patterns. Finally, there should be a phase characteristic of the large $\theta$ regime in which mainly the intermediate states with $S_i=\frac{1}{3}{\rm sgn}(\xi_i)$ are activated leading to a performance with $m=2/5$, $q=1/9$ and again $d_H=2/9$. This is the phase $FM_{\Delta}$, which should have an overlap with the phase $FM_1$ at intermediate $\theta$. These expectations have been confirmed by means of the solutions to the saddle-point equations in the $\alpha\rightarrow 0$ limit and some of the results can be found in earlier work \cite{BRS94}. The four regimes are given by $\theta <1/4$,	
${1/4}<\theta <{3\Delta/4}$, ${3\Delta/4}<\theta<3/4$ and ${3/4}<\theta$.
They do not depend on the connectivity and are therefore independent of the architecture of the network, in accordance with earlier results either on the fully connected or the extreme symmetrically dilute network \cite{BRS94,BCS00}.     

On the other hand, we confirm the symmetry of the limiting $\alpha_c$ for $\theta\rightarrow\infty$ and for $\theta = 0$, in accordance with earlier results \cite{BRS94}. We also find that the optimal performance line appears within the $FM_1$ phase and that the network has a relatively high performance with a small Hamming distance in that phase, with an overlap at the critical phase boundary that is $0.8$ of that at $\alpha=0$. 

When the connectivity is reduced to $c\simeq 0.48$, the distinction between the phases $FM_{3\Delta}$ and $FM_1$ starts to disappear, as shown by the enlarged gap in Fig. 3(b), allowing for a continuous change into the high performance phase for intermediate $\alpha$. Note that there is still a discontinuous phase boundary between the phases $FM_{3\Delta}$ and $FM_{\Delta}$ and that the presence of this phase boundary is important in order to inhibit the transition to the low performance phase $FM_{\Delta}$. 

Furthermore, we still find four regimes for low $\alpha$ and that the three main retrieval phases, $FM_{3\Delta}$, $FM_1$ and $FM_{\Delta}$, end discontinuously at the critical phase boundary $\alpha_c$. The optimal performance line is still purely within the $FM_1$ phase, as in the previous case, and the network has a high performance up to $\alpha_c$, with an overlap close to one on the phase boundary, for $\theta$ around $0.5$.

As the connectivity is further decreased to $c=0.1$ we find the phase diagram shown in Fig. 3(c) with the three main retrieval phases that disappear discontinuously at $\alpha_c$ and the four low-$\alpha$ regimes discussed above. The distinction between the phases $FM_{3\Delta}$ and $FM_1$ disappears now at lower $\alpha$ and the optimal network performance in the central phase $FM_1$ can be reached continuously within a considerable range of $\alpha$ from the $FM_{3\Delta}$ phase. As in the previous cases, there is a coexistence region between the phases $FM_{3\Delta}$ and $FM_1$, now only for small $\alpha$. Moreover, there is no need now for a specific choice of threshold parameter $\theta$ in order to access most of the high performance domain of the network. Incidentally, note that the continuous retrieval phase boundary for the common phase $FM_{3\Delta}$ and $FM_1$ is similar to that found for the fully connected network with pattern activity $a=7/9$ \cite{BRS94}. We have no further insight, at present, of this feature.

Finally, in order to check the overall simplification of the phase diagrams that appears with decreasing connectivity, we also present results for $c=0.01$ that are shown in Fig. 3(d). The phase boundaries are still lines of discontinuous transitions and the distinction between the four regimes is restricted to even lower values of $\alpha$. There is now a considerably larger region of continuous changeover from the phase $FM_{3\Delta}$ to the phase $FM_1$, with access to optimal performance, without the need of a fine adjustment in $\theta$.   

\subsection{The continuous response network}

In the case of $Q=\infty$ we again consider uniformly distributed patterns between $-1$ and $1$, implying that $a=1/3$, and restrict the results to the zero temperature case. The discontinuous transitions to the ordered phase are shown in full lines in the $\alpha-\theta$ phase diagram for decreasing connectivity in Fig. 4, where we omit again the thermodynamic transitions and the long-dashed lines indicate now the onset of the binary-network behavior. Note that the disappearance of the ordered phase takes place at $\theta=1/2$ for any finite connectivity, as in the case of the fully connected network and in networks of different architecture, like the extremely asymmetric diluted and the $Q$-Ising layered network \cite{BRS94,BSVZ94,BSV94}. In the case of the extremely diluted networks, the retrieval phase boundaries have a reentrance for $\theta\geq \frac{1}{2}$ \cite{BCS00}. This seems to be a feature of the $c\rightarrow 0$ limit, as one can see from our further results for the connectivity dependence of the maximum storage capacity $\alpha_{max}$ and the corresponding $\theta_{max}$, both shown in Fig. 5.

As in the case of both the fully connected and the symmetrical extremely diluted network, and in contrast with the $Q=3$ state network, we find that even at zero temperature most of the retrieval regions for different $c$ are stable to replica-symmetry-breaking perturbations, that is for $\theta$ above the AT lines. This includes the maximum storage capacity  and it follows from a positive replicon eigenvalue for this case obtained from Eq.(38),
\begin{equation}
\lambda_R=1-\frac{{\alpha}rc}{4q{\tilde\theta}^2}\left\langle\int_{|m\xi+\sqrt{\alpha rc}z|\leq 2\tilde\theta}{\cal D}z\,
        \right\rangle_{\{\xi^{\nu}\}}=1-\frac{\alpha rc\chi}{2q\tilde\theta}\,,
\label{40}
\end{equation}
where $\chi$ is the susceptibility for the continuous network presented in the Appendix. The AT line is given by $\lambda_R=0$ and, again, in both the $c=1$ and the $c\rightarrow 0$ limit, in which $rc=q$, this result coincides with that for the fully connected and the symmetrical extremely diluted network \cite{BRS94,BCS00}.

\section{Summary and concluding remarks}

We derived in this work the replica symmetric mean-field theory for $Q$-Ising attractor networks with low-activity patterns and arbitrary symmetric dilution of the synaptic connections. We extended earlier studies on the retrieval behavior and thermodynamic properties of either fully connected or symmetrical extremely diluted $Q$-Ising neural networks with low activity patterns, in order to study the effects of a gradual dilution of the synaptic connections guided by the motivation that neurons in biological networks of associative memory are neither fully connected nor very sparsely linked to other neurons. We are mainly interested in the nature of the phase transitions to locally stable retrieval states and in the role that synaptic dilution has in either reducing or destroying sharp transitions motivated by the plasticity of biological networks. In the context of the networks studied here, we focus mainly on the dependence of the retrieval properties on the threshold or gain parameter $\theta$ for decreasing connectivity $c$ and ask to which extent can a network go from a given locally stable retrieval state to a nearby high-performance state without crossing phase boundaries of discontinuous transitions.

In order to answer that question, one has to look for appropriate phase diagrams that were obtained here in replica-symmetric mean-field theory. Since deviations from that theory are very small and appear only in a small region near $T=0$, we
may still draw relevant conclusions from those diagrams. The explicit phase diagrams obtained in this work apply to uniformly distributed patterns and to networks with arbitrary symmetric dilution. The behavior of both the fully connected and the extremely diluted network are recovered when $c=1$ and $c\rightarrow 0$, respectively, in that our general saddle-point equations become identical to those for either case that have been obtained before \cite{BRS94,BCS00}. 

We find that common features of the $\alpha\rightarrow 0$ limiting behavior in either fully connected or symmetrical extremely diluted networks also appear for arbitrary finite connectivity $c$. This confirms the expectations of earlier works that pointed out the architecture independent nature of some properties \cite{BRS94,BCS00}. Among these is the particular $\theta=0.5$ where the thermodynamic transition ends in the $Q=3$ state network and where the optimal performance takes place for low $\alpha$, both for $T=0$. The common limiting $T$ in the $(\alpha,T)$ phase diagram for varying $c$ is a further property of this kind, as well as the four distinct domains in the $Q=4$ state network and the $\theta=0.5$ limiting threshold for the $Q=\infty$ network at $\alpha=0$.

The main dependence of the behavior of the network on the connectivity arises for finite $\alpha$. For both odd and even $Q$, we find that a common feature that appears with an increase in the dilution of the synaptic connections is to suppress selected sharp phase boundaries of discontinuous transitions that make the optimal performance domain readily accessible to a wide region of low-threshold locally stable retrieval states. Note that, on the other hand, the sharp boundary for the $Q=4$ state network between the low performance phases $FM_{3\Delta}$ and $FM_{\Delta}$ survives synaptic dilution, at least to quite an extent. These features of the network for small but finite connectivity appear long before the extremely diluted limit and they should be of considerable interest.

Concentrating, for simplicity, on the $Q=3$ state network we also found that the boundaries between thermodynamic transitions are suppressed by an increase of the synaptic dilution, and expect a similar behavior for the $Q=4$ state network.  

The results of our work may be used to infer the behavior of other networks.  Since the fully connected network is strongly sensitive to pattern activity, one may consider other than uniformly distributed patterns \cite{BRS94}. There are, essentially, two kinds of phase diagrams in that case about which we can make definite predictions. One is the type of phase diagram for patterns of relatively large activity that has mostly a decreasing phase boundary with increasing $\theta$ and an optimal performance line that appears only at small $\theta$. The other type, that appears for medium or small pattern activity, has an optimal performance only at intermediate $\theta$, like the cases shown in Fig. 3, and both types appear for $Q=3$ and $Q=4$, while only the first type seems to appear for $Q=\infty$. We expect that the main effect of a finite synaptic dilution on the first type of phase diagram is simply to shift the retrieval phase boundaries upwards towards a larger $\alpha_c$. In the second type of phase diagrams, however, we expect also a disappearance of the discontinuous phase boundary between the $FM_{3\Delta}$ and $FM_1$ phases, in essentially the same way we found in the present work, allowing for a smooth changeover from states of non-optimal to those of optimal performance.

The main result of this work, that partially connected multi-state Hebbian networks can attain near-optimal performance without a fine tuning of neuron activity may be a simplified statistical mechanics explanation of why biological memory networks seem to prefer low-activity patterns between partially connected neurons. Of course, biological networks have asymmetric synaptic connections which may lead through a dynamic evolution to different stationary states, the search of which is certainly an interesting issue that deserves a separate investigation, currently in progress.

The study of the effects of symmetric synaptic dilution may be extended to other problems that deal with associative memory, like the categorization problem as a classification task in $Q$-Ising networks \cite{ETD99}. This has been done recently for $Q=2$ \cite{KT99} and there is work in progress for general $Q$ \cite{ET01}.

{\bf Acknowledgments}

We thank D. Boll\'e and G. M. Shim for clarifying a point on a previous version of the manuscript. This work was financially supported in part by CNPq 
(Conselho Nacional de Desenvolvimento Cient{\'\i}fico e Tecnol{\'o}gico)
Brazil.

\newpage
\section{Appendix}

We present here, for completeness, the explicit expressions for the saddle-point equations obtained for the symmetrically diluted network with arbitrary connectivity $c$ and uniformly distributed patterns, for $Q=3$, $Q=4$ and $Q=\infty$. 

For $Q=3$ taking patterns $\pm 1$ with probability $a/2$ and $0$ with probability $(1-a)$ we have
\begin{equation}
m=\int{\cal 
	D}z\,S_{\beta}(m+\sqrt{{\alpha}rc}z,\tilde\theta)\,,
\label{A.1}
\end{equation}
\begin{equation}
q=\int{\cal D}z\,\left[aS^{2}_{\beta}(m+\sqrt{{\alpha}rc}z,\tilde\theta)
            +(1-a)S^{2}_{\beta}(\sqrt{{\alpha}rc}z,\tilde\theta)
            \right]\,,
\label{A.2}
\end{equation}
\begin{equation}
\chi=\frac{1}{\sqrt{\alpha rc}}\int{\cal D}z\,
	z\left[aS_{\beta}(m+\sqrt{{\alpha}rc}z,\tilde\theta)
            +(1-a)S_{\beta}(\sqrt{{\alpha}rc}z,\tilde\theta)
            \right]\,,
\label{A.3}
\end{equation}
with $S_{\beta}(h,\tilde\theta)$ given by Eq.(39). These equations reduce to the equations for the extreme symmetrically diluted network, as well as for the fully connected network, in the $c\rightarrow 0$ limit and $c=1$, respectively. The same applies for the cases of $Q=4$ and $Q=\infty$, presented below. 

In the zero-temperature limit ($\beta\rightarrow\infty$), the integrations over the Gaussian variable $z$ can be done explicitly. In the $Q=3$ case we obtain, for $\tilde\theta\geq 0$,
\begin{equation}
m=\frac{1}{2}\left[{\rm erf}(\frac{m+\tilde\theta}{\sqrt{2\alpha rc}})                             
+{\rm erf}(\frac{m-\tilde\theta}{\sqrt{2\alpha rc}})\right]\,,
\label{A.4}
\end{equation}
\begin{equation}
q=1-\frac{a}{2}\left[{\rm erf}(\frac{m+\tilde\theta}{\sqrt{2\alpha rc}})                             
-{\rm erf}(\frac{m-\tilde\theta}{\sqrt{2\alpha rc}})+(1-a){\rm erf}(\frac{\tilde\theta}{\sqrt{2\alpha rc}})\right]\,,
\label{A.5}
\end{equation}
\begin{equation}
\chi=\sqrt{\frac{1}{2\pi\alpha rc}}\left[a\exp(-\frac{(m+\tilde\theta)^2}
         {2\alpha rc})                             
         +a\exp(-\frac{(m-\tilde\theta)^2}{2\alpha rc})
         +2(1-a)\exp(-\frac{\tilde\theta^2}{2\alpha rc})\right]\,,
\label{A.6}
\end{equation}
with the relation between $r$, $q$ and $\chi$ given by Eq.(36).

For the $Q=4$ state model with states and uniformly distributed patterns 
that take the values $-1$, $-1/3$, $+1/3$ and $+1$, $a=5/9$ and we obtain
\begin{equation}
m=\frac{9}{10}\int{\cal 
	D}z\,\left[S_{\beta}(m+\sqrt{{\alpha}rc}z,\tilde\theta)
+S_{\beta}(\frac{m}{3}+\sqrt{{\alpha}rc}z,\tilde\theta)\right]\,,
\label{A.7}
\end{equation}
\begin{equation}
q=\frac{1}{2}\int{\cal D}z\,\left[S^{2}_{\beta}(m+\sqrt{{\alpha}rc}z,\tilde\theta)
            +S^{2}_{\beta}(\frac{m}{3}+\sqrt{{\alpha}rc}z,\tilde\theta)
            \right]\,,
\label{A.8}
\end{equation}
\begin{equation}
\chi=\frac{1}{2\sqrt{\alpha rc}}\int{\cal D}z\,
	z\left[S_{\beta}(m+\sqrt{{\alpha}rc}z,\tilde\theta)
            +S_{\beta}(\frac{m}{3}+\sqrt{{\alpha}rc}z,\tilde\theta)
            \right]\,.
\label{A.9}
\end{equation}
where
\begin{equation}
S_{\beta}(h, \tilde\theta)=\frac{\sinh(\beta h)
+\frac{1}{3}\exp({8\beta \tilde\theta}/9)\sinh({\beta h}/3)}
{\cosh(\beta h)+\exp({8\beta \tilde\theta}/9)\cosh({\beta h}/3)}\,.
\label{A.10}
\end{equation}
In the zero-temperature limit
\begin{equation} 
S_{\infty}(h, \tilde\theta)=sign(h)\left[1+2\theta(|h|
-4\tilde\theta /3)\right]\,,
\label{A.11}
\end{equation}
and the above equations yield, for positive $\tilde\theta$,
\begin{eqnarray}
m&=&\frac{3}{10}\left[{\rm erf}(\frac{3m+4\tilde\theta}
     {3\sqrt{2\alpha rc}})
     +{\rm erf}(\frac{3m-4\tilde\theta}{3\sqrt{2\alpha rc}})
     +{\rm erf}(\frac{m}{\sqrt{2\alpha rc}})\right]\nonumber\\
     &+&\frac{1}{10}\left[{\rm erf}(\frac{m+4\tilde\theta}
     {3\sqrt{2\alpha rc}})                             
     +{\rm erf}(\frac{m-4\tilde\theta}{3\sqrt{2\alpha rc}})
     +{\rm erf}(\frac{m}{3\sqrt{2\alpha rc}})\right]\,,
\label{A.12}
\end{eqnarray}
\begin{equation}
q=1-\frac{2}{9}\left[{\rm erf}(\frac{3m+4\tilde\theta}
{3\sqrt{2\alpha rc}})                             
-{\rm erf}(\frac{3m-4\tilde\theta}{3\sqrt{2\alpha rc}})
+{\rm erf}(\frac{m+4\tilde\theta}{3\sqrt{2\alpha rc}})                             
-{\rm erf}(\frac{m-4\tilde\theta}{3\sqrt{2\alpha rc}})\right]\,,
\label{A.13}
\end{equation}
\begin{eqnarray}
\chi&=&\frac{1}{3\sqrt{2\pi\alpha rc}}
         \left[\exp(-\frac{(3m+4\tilde\theta)^{2}}
          {18\alpha rc})                             
         +\exp(-\frac{(3m-4\tilde\theta)^{2}}{18\alpha rc})
         +\exp(-\frac{m^{2}}{2\alpha rc})\right.\nonumber\\
    &+&\left.\exp(-\frac{(m+4\tilde\theta)^{2}}{18\alpha rc})                             
         +\exp(-\frac{(m-4\tilde\theta)^{2}}{18\alpha rc})
         +\exp(-\frac{m^{2}}{18\alpha rc})\right]\,.
\label{A.14}
\end{eqnarray}

Finally, in the zero-temperature limit for $Q=\infty$ and uniformly distributed patterns between $-1$ and $1$, implying $a=1/3$, we obtain for $\tilde\theta\geq 0$, 
\begin{equation}
m=\frac{3}{2}\int_{-1}^{+1}{\rm d}\xi\,
{\xi}\left[(1+\frac{m\xi}{2\tilde\theta}){\rm erf}({\rm B}(m))
+\frac{1}{\tilde\theta}\sqrt{\frac{\alpha rc}{2\pi}}
\exp(-{\rm B}^{2}(m))\right]\,,
\label{A.11}
\end{equation}
\begin{eqnarray}
q=1&+&\frac{1}{2}\int_{-1}^{+1}{\rm d}\xi\,
       \left[(\frac{\alpha rc+(m\xi)^{2}}{(2\tilde\theta)^{2}}-1)
       {\rm erf}({\rm B}(m))\right.\nonumber\\
      & &\quad\quad\quad+\left.\frac{1}{\tilde\theta}\sqrt{\frac{\alpha rc}{2\pi}}
       (\frac{m\xi}{2\tilde\theta}-1)\exp(-{\rm B}^{2}(m))\right]\,,
\label{A.12}
\end{eqnarray}
\begin{equation}
\chi=\frac{1}{4\tilde\theta}\int_{-1}^{+1}{\rm d}\xi\,
{\rm erf}({\rm B}(m))\,.
\label{A.13}
\end{equation}
where
\begin{equation}
{\rm B}(m)=\frac{2\tilde\theta+m\xi}{\sqrt{2\alpha rc}}\,,
\label{A.14}
\end{equation}
and this integrations can be performed directly.

\newpage
\noindent
{\bf Figure Captions}\\

\noindent
FIG. 1. Phase diagram $(\alpha,\theta)$ for the $Q=3$ state network with
uniformly distributed patterns and connectivity $c$ as shown, at $T=0$. The
full and the long-dashed lines represent the retrieval and the thermodynamic
transition, respectively. The latter ends on the right at the (dotted) optimal
performance line and the spin-glass phase appears at the left of the
paramagnetic phase boundary indicated by the short-dashed lines. The two
retrieval regions I and II are discussed in the text and the inset corresponds to $c=0.63$ where the distinction betweeen these phases starts to disappear.

\noindent
FIG. 2. Stable phase diagram to the right of the de Almeida-Thouless
(dot-dashed) line in the $(\alpha,T)$ plane for the $Q=3$ state network with
uniformly distributed patterns, $\theta=0.2$ and connectivity $c$ as
shown. The full lines represent the retrieval transition.

\noindent
FIG. 3. Phase diagram $(\alpha,\theta)$ for the $Q=4$ state
network with 
uniformly distributed patterns and connectivity (a)$c=0.5$, (b)$c=0.48$,
(c)$c=0.1$ and (d)$c=0.01$, at $T=0$. The full lines represent retrieval
transitions and the optimal performance is indicated in dotted lines. The
central region is the best performance phase $FM_1$ and there is a low-$\alpha$
coexistence region between phases $FM_{3\Delta}$ and $FM_1$.
 

\noindent
FIG. 4. Phase diagram $(\alpha,\theta)$ for the $Q=\infty$ state network with
uniformly distributed patterns and connectivity $c$ as shown, at
$T=0$. The full lines represent retrieval transitions, the optimal performance
is indicated in dotted lines and the onset of the binary network is shown by
long-dashed lines. The phases are stable to the right of the de Almeida-Thouless
line (dot-dashed).

\noindent
FIG. 5: Connectivity dependence for the maximum storage capacity
$\alpha_{max}$ and the corresponding $\theta_{max}$ in the
$Q=\infty$ state network with uniformly distributed patterns at $T=0$.

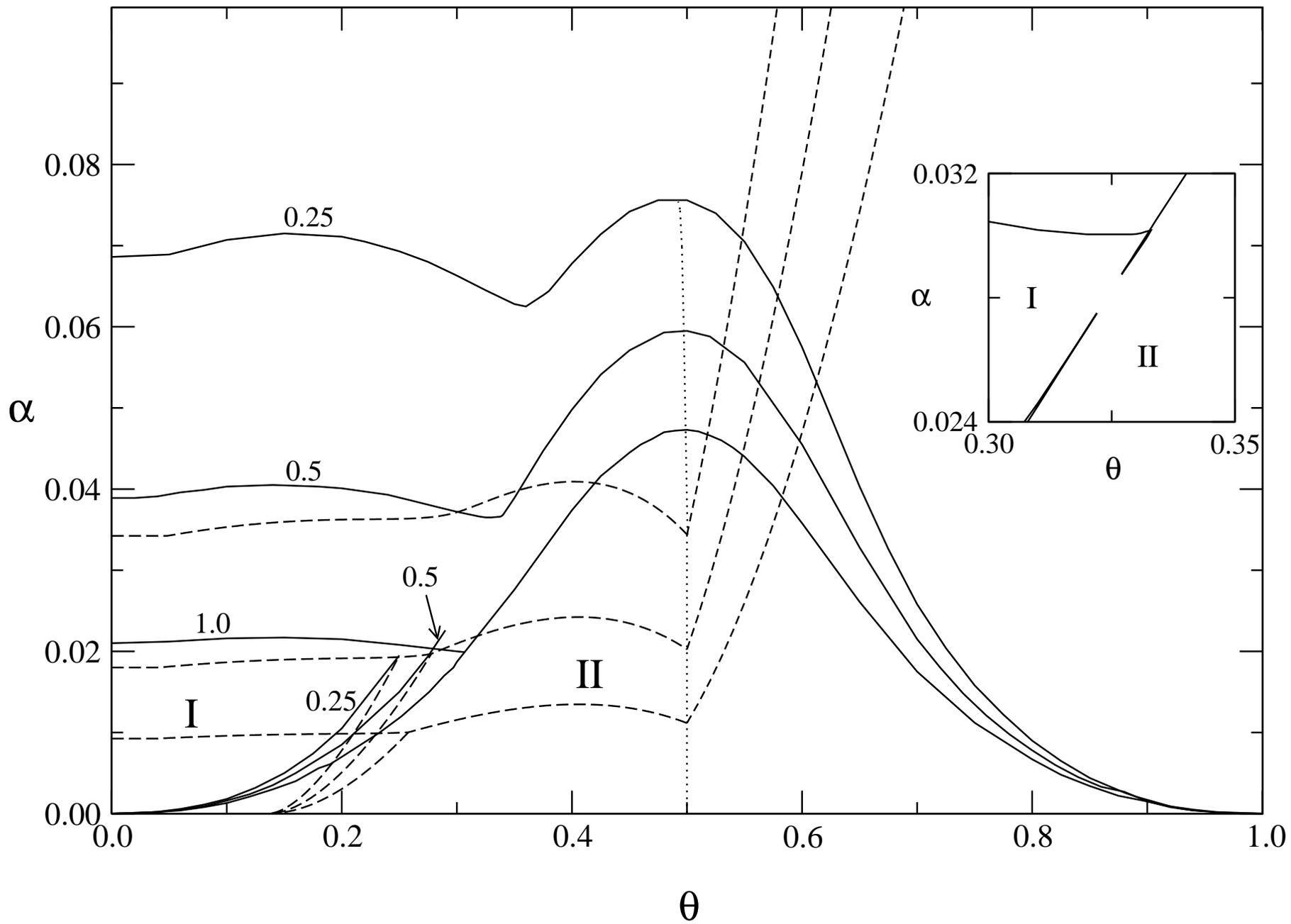
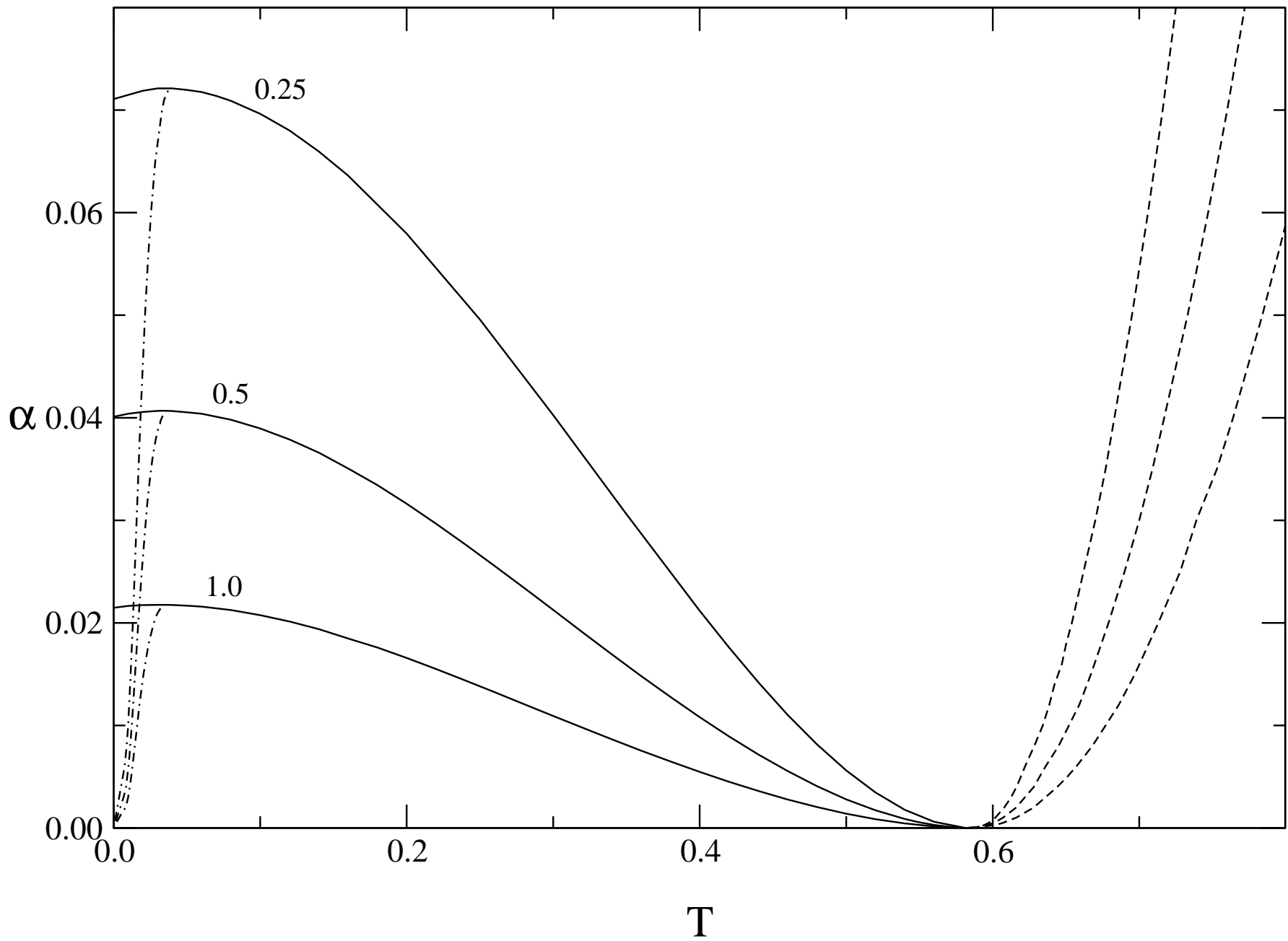
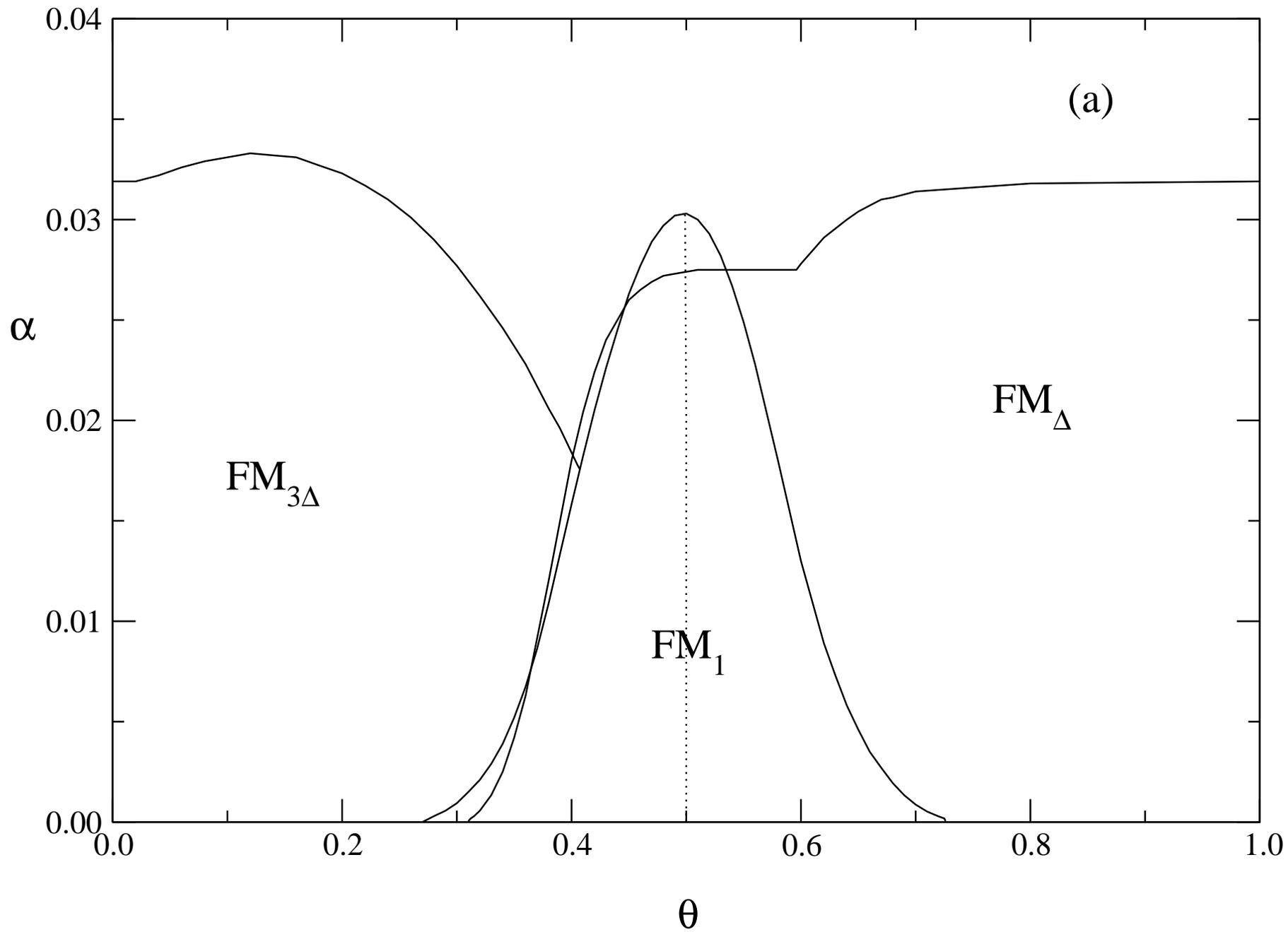
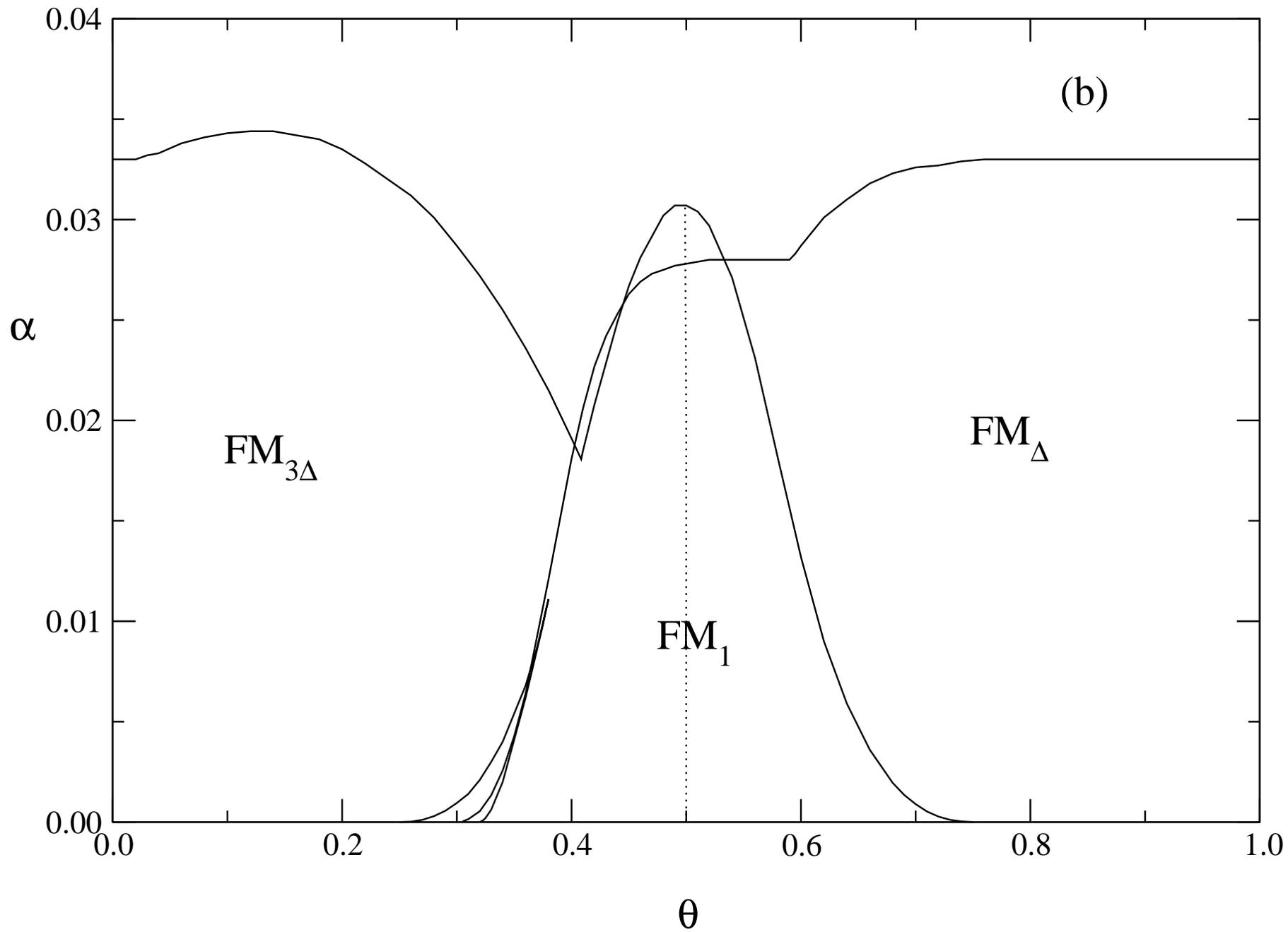
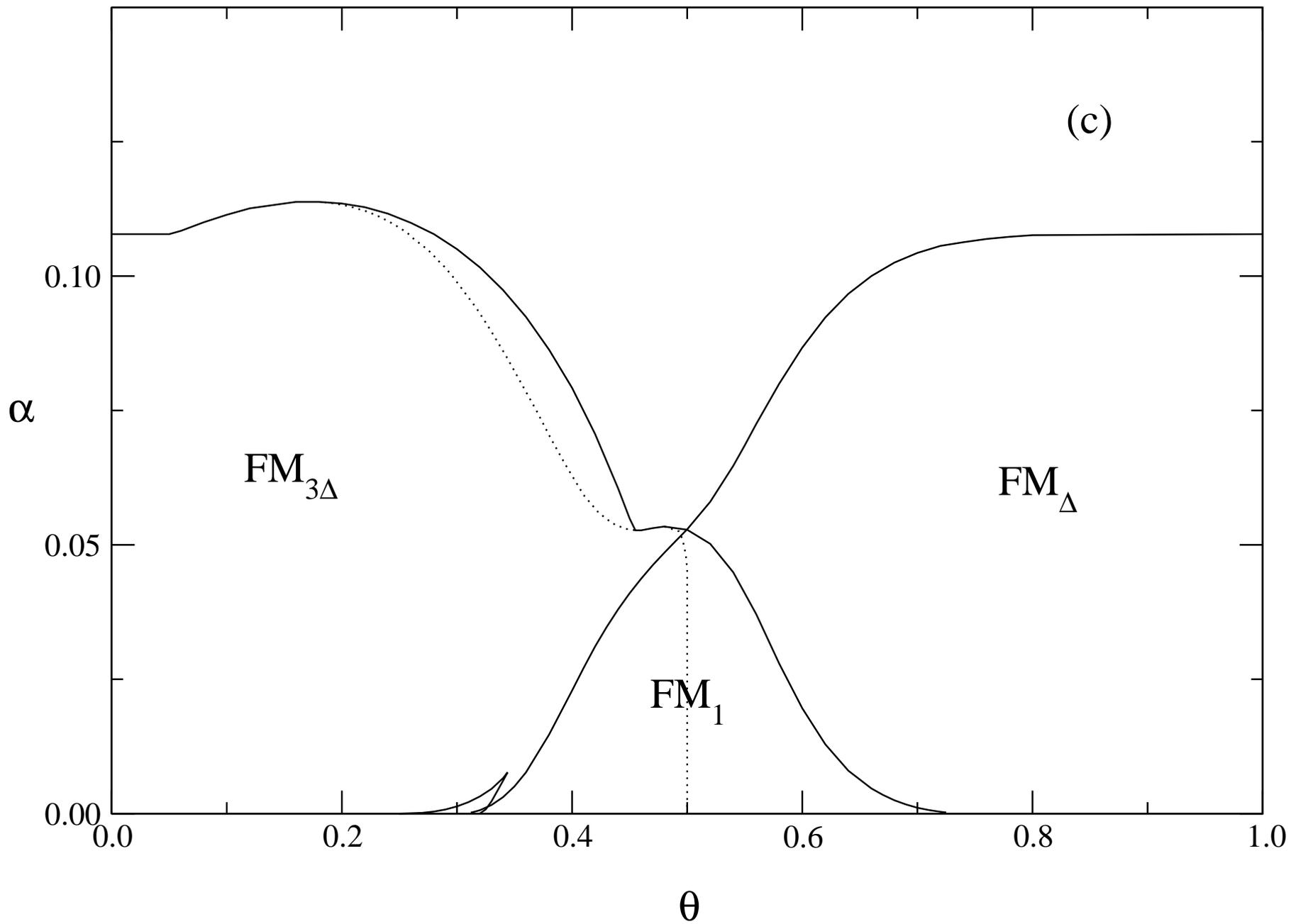
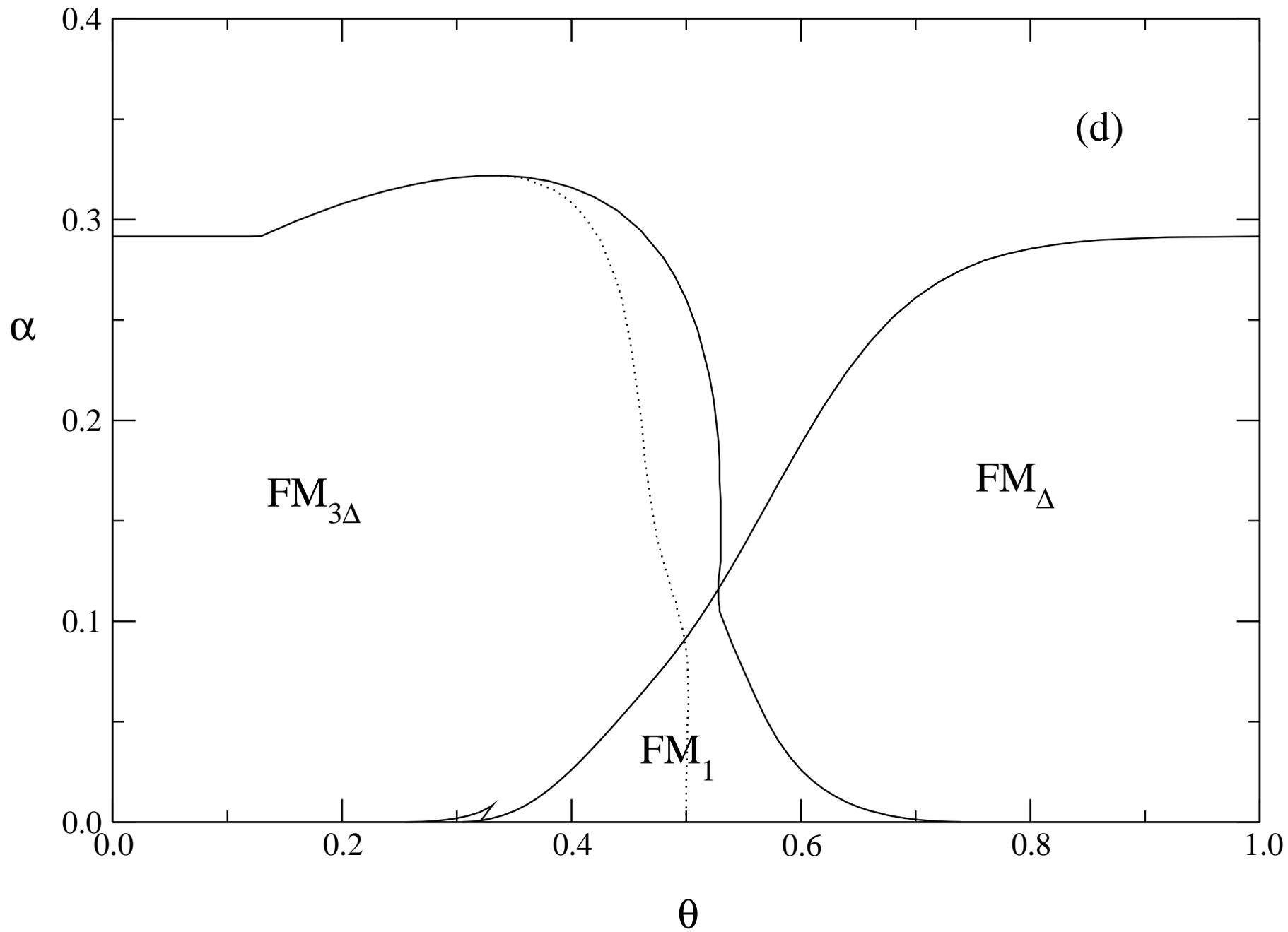
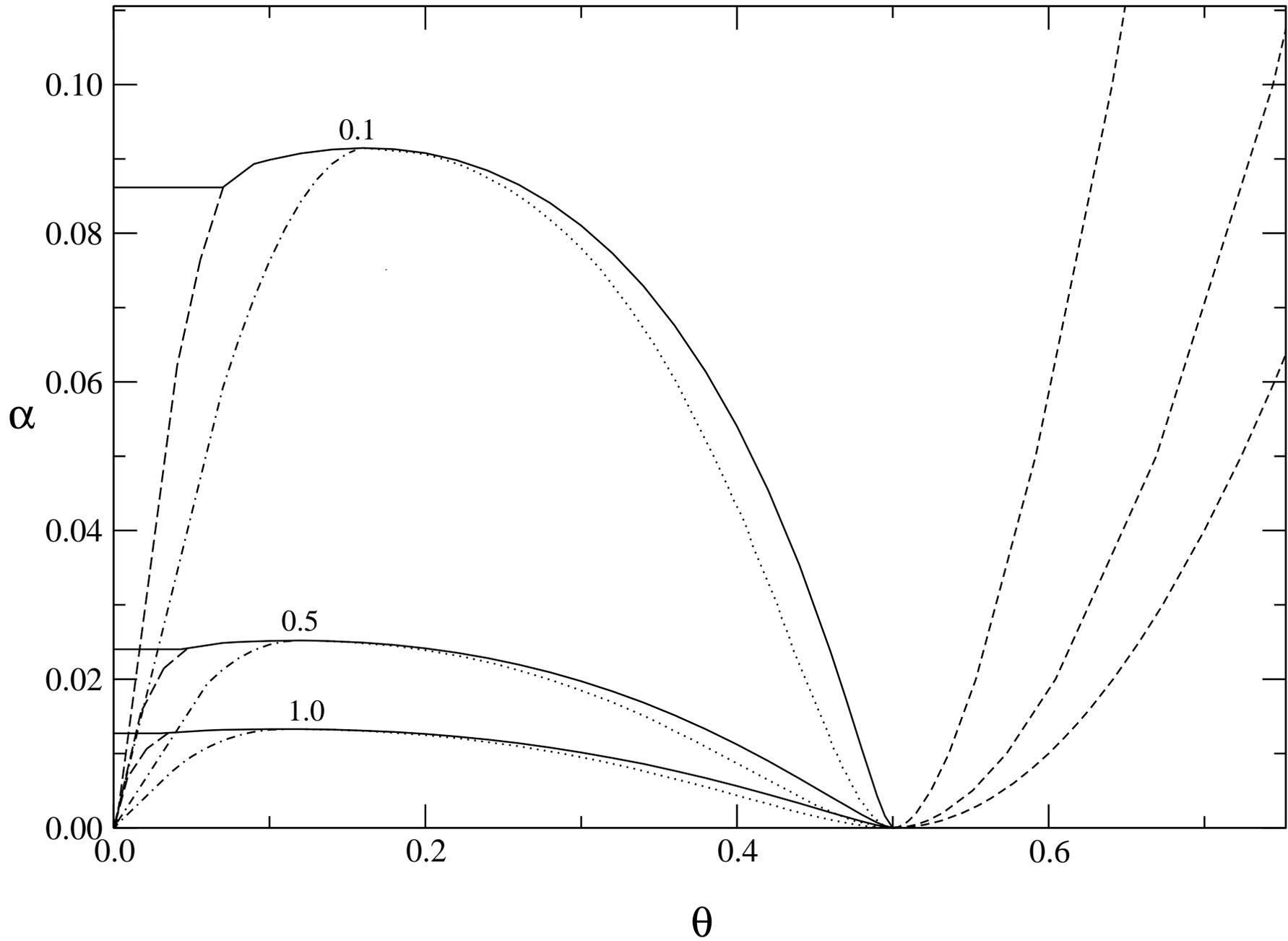
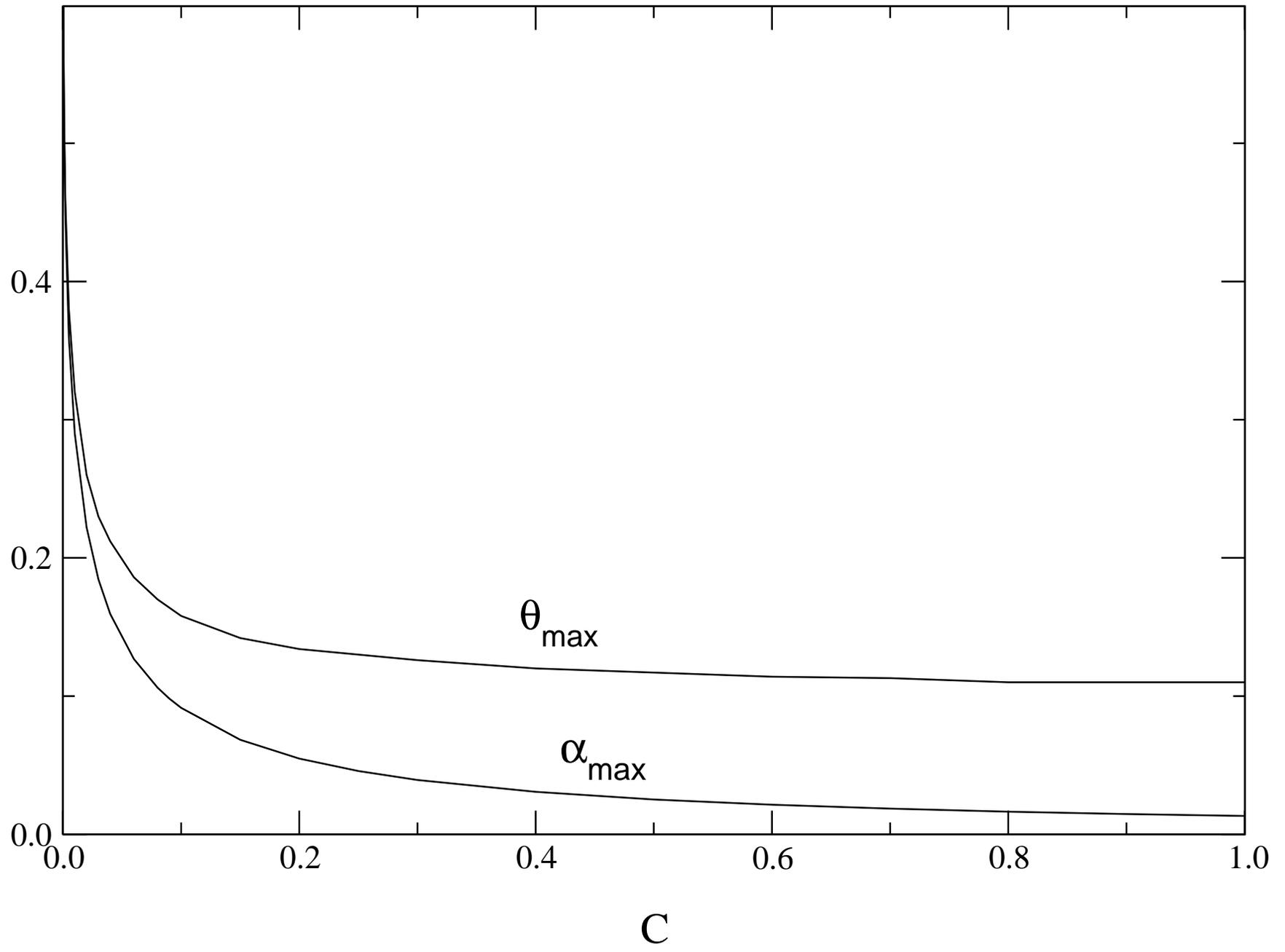
\end{document}